\documentclass[twocolumn]{article}
\pdfoutput=1
\usepackage{arxiv}

\usepackage[utf8]{inputenc}
\usepackage[T1]{fontenc}
\usepackage{amsmath}
\usepackage{amssymb}
\usepackage{graphicx}
\usepackage{hyperref}
\usepackage{url}            
\usepackage{booktabs}       
\usepackage{amsfonts}       
\usepackage{nicefrac}       
\usepackage{microtype}      
\usepackage{xcolor}         
\usepackage{enumitem}
\usepackage{abstract}
\usepackage{listings}
\usepackage{makecell}
\usepackage{tikz}
\usetikzlibrary{shapes.geometric,arrows,positioning}
\usetikzlibrary{shapes.symbols}
\usepackage{pgfplots}
\usepackage{pgfplotstable}
\usepgfplotslibrary{statistics}
\usepgfplotslibrary{fillbetween}
\usepackage{array}
\usepackage{colortbl}
\pgfplotsset{compat=newest}
\usepackage[section]{placeins}

\definecolor{lightgray}{RGB}{240,240,240}
\definecolor{lightblue}{RGB}{230,240,255}

\title{Patience is all you need! An agentic system for performing scientific literature review}

\date{\today}

\usepackage[affil-it]{authblk}

\author[1]{David Brett
\thanks{\texttt{david.brett@recursion.com}}}
\author[1]{Anniek Myatt
\thanks{\texttt{anniek.myatt@recursion.com}}}
\affil[1]{Recursion Pharmaceutical, Salt Lake City, UT 84101, USA}

\hypersetup{
pdftitle={Patience is all you need! An agentic system for generating scientific literature reviews},
pdfauthor={David~Brett, Anniek~Myatt},
pdfkeywords={LLMs,Q\&A,RAG,information retrieval,agents},
pdflang={en-GB}
}

\begin{document}

\twocolumn[
\maketitle
\begin{abstract}
Large language models (LLMs) have grown in their usage to provide support for question answering across numerous disciplines. 
The models on their own have already shown promise for answering basic questions, however fail quickly where expert domain knowledge is required or the question is nuanced. 
Scientific research often involves searching for relevant literature, distilling pertinent information from that literature and analysing how the findings support or contradict one another. 
The information is often encapsulated in the full text body of research articles, rather than just in the abstracts. Statements within these articles frequently require the wider article context to be fully understood. 
We have built an LLM-based system that performs such search and distillation of information encapsulated in scientific literature, and we evaluate our keyword based search and information distillation system against a set of biology related questions from previously released literature benchmarks. 
We demonstrate sparse retrieval methods exhibit results close to state of the art without the need for dense retrieval, with its associated infrastructure and complexity overhead.
We also show how to increase the coverage of relevant documents for literature review generation.
\end{abstract}
\vspace{2cm}
]
\saythanks{}
\section{Introduction}
The use of large language models (LLMs) has grown dramatically in recent years with foundation models displaying impressive capabilities across many applications~\cite{10.48550/arxiv.2108.07258,10.48550/arxiv.2303.12712}, particularly in the context of agentic workflows for the orchestration of tools ~\cite{10.1038/s41586-023-06792-0, bran2023chemcrowaugmentinglargelanguagemodels}.
One of these applications is to make use of LLM augmented agents for search and distillation of information for question answering (Q\&A) in the biomedical domain~\cite{lala2023paperqa,10.48550/arxiv.2408.01107}. 
The focus of building LLM applications for the biomedical domain is largely driven by free data availability via PubMed (titles and abstracts) and PubMed Central (full text journals), where as licensing terms for equivalent data sources for other domains, e.g. chemistry, are more restrictive.

Including explicit information from data sources such as scientific literature in the LLM context for answering expert-level questions is crucial to reduce the likelihood of getting wrong answers ("hallucinations"), particularly when specific details from complex scientific texts have to be considered that will not have appeared frequently in the LLMs training data. 
In addition, domain experts only have confidence in an answer if source attributions, in the form of literature references, are provided alongside the answer. This renders answers of raw LLMs useless, even if they are correct.

Retrieval augmented generation (RAG)~\cite{NEURIPS2020_6b493230} has been widely employed to overcome these issues~\cite{gao2024retrievalaugmentedgenerationlargelanguage}. The approaches to retrieval and information distillation have to be tailored specifically to the characteristics of the data sources and problem statement.
Recently FutureHouse, a company specialising in AI-powered research tools, released a system for question answering on scientific literature using LLMs, LitQA2 ~\cite{10.48550/arxiv.2312.07559,paperqa2}, alongside a benchmark dataset to evaluate its performance. 

\subsection{Limitations of existing benchmarks}
While benchmarks for biomedical question and answering have existed for a while, they typically evaluated a language model's ability to answer multiple choice questions given a text extract (e.g PubMedQA~\cite{10.48550/arxiv.1909.06146}). 
Such benchmarks lack the scope to assess the retrieval capabilities of LLM augmented systems. The LitQA2 benchmark~\cite{10.48550/arxiv.2407.10362} has been developed to support evaluation of Q\&A on literature, testing both the retrieval and information extraction abilities of LLM augmented agents. 
It is still vulnerable to the issues presented by multiple-choice based benchmarks~\cite{10.48550/arxiv.2403.17752,10.48550/arxiv.2406.07545} and is necessarily a limited measure for the potential effectiveness of a system for real world applications where long form answers are commonly required.

Typically in scientific research questions tend to be open ended and reading a wide selection of literature as part of a review is more important than finding a specific piece of information contained in a single paper, which is what benchmarks such as PubMedQA and LitAQ2 are designed to evaluate. 
An inherent difficulty with considering the full text of articles in a benchmark is the licensing of those articles. For example not all the of papers used by LitQA2 are openly licensed for machine reading which technically prevents reproducing the results without paying for the papers. 

Existing benchmarks do not attempt to consider the coverage of literature used to answer longer form answers/review generation. Typical review papers might have around 100 references as part of answering a question while benchmarks to date are typically answered with a single paper.  
In this paper we present a benchmark of retrieval coverage not just for a single paper but for the set used within actual review articles.

\subsection{Agentic systems for retrieval}
We define an agentic system as one which is made up of multiple LLM-augmented agents which are orchestrated together to perform a task.
There are a number of scientific literature agentic systems for different tasks\cite{Glickman2024AI,paperqa2}.

Many retrieval systems perform a two stage retrieval with an initial search and then re-ranking of the results. 
In both stages dense retrieval (in which the whole document is encoded as a dense vector)~\cite{neelakantan2022textcodeembeddingscontrastive,douze2024faisslibrary} and sparse retrieval (in which the vocabulary of the document is encoded)~\cite{10.1561/1500000019,formal2021spladesparselexicalexpansion} might be used.

Considering the first stage of retrieval, a limitation of dense retrieval lies in its sensitivity to the chunk size of the embedded text as well as sensitivity to the choice of encoder.
With respect to the former, the correlation of a mention of a term in one section of a paper with a mention in another section might be missed if the sections have been turned into separate text chunks and corresponding embedding vectors. 
Furthermore for large document corpus it can be expensive to index for dense retrieval.

Sparse retrieval on the other hand lacks the semantic meaning between terms in the vocab of a document. Furthermore it doesn't explicitly handle the issue of synonyms, which query expansion aims to address~\cite{https://doi.org/10.1155/2014/132158}. 
Dense retrieval effectively performs this query expansion within in the embedding of the model used for encoding the text. 
In our system we make use of the LLM to perform the query expansion not just in terms of adding synonyms but also in understanding what vocabulary terms might be necessary in a document to find the answer, rather than just expanding on the question. 

Recent studies such as LongRAG~\cite{jiang2024longragenhancingretrievalaugmentedgeneration} have demonstrated that the original techniques for retrieval augmented generation to chunk text into short segments for both retrieval and to provide as prompts for zero shot Q\&A~\cite{NEURIPS2020_6b493230} are often no longer necessary given the increasing context size of models (for example Claude 3.5 Sonnet has a 200K token context length~\cite{Claude3}). 
Previously, handling long contexts was seen to be an issue for models such as GPT 3.5, which struggled with information retrieval from extended passages~\cite{liu-etal-2024-lost}.
Retrieval across long contexts have been demonstrated to be fairly reliable ~\cite{needleinahaystack,Claude3}, however the context size is still finite so putting 20-30 complete full text scientific articles in verbatim is still not feasible.
This motivates article chunking and re-ranking based on relevance to answering the question for our use case of literature review generation.

The re-ranking of results is typically done using dense retrieval with powerful techniques such as cross-encoders~\cite{ANNAMORADNEJAD2024123685, patel2024retrieveexplainevidencedrivenpredictions} where a model encodes the semantic similarity of the question to the document. 
In our system we introduce a re-ranking based on sparse retrieval using BM25L~\cite{10.1145/2009916.2010070} combining the ranks of multiple queries with expanded synonyms generated by an LLM.

In a further stage for retrieval in our system we let the LLM decide whether to keep or reject a document for use in answering the question, and also summarising the document in context. A similar approach to this is used by PaperQA2~\cite{paperqa2}.

Our agentic system for retrieval differs to many existing agentic systems in that it is solely based on sparse retrieval with query expansion performed by LLMs rather than making use of any dense retrieval techniques.
This is chosen to enable better use of the large context sizes available in newer models~\cite{Claude3}.

In addition, we introduce a method to expand the relevant literature that is included in an LLM-generated review. 
This approach involves generating a series of expansive questions based on the initial answer from a retrieval-and-answering round. 
This helps to uncover more detailed information and identify potentially contradicting research, which is particularly important for literature review generation, where researchers synthesise a broad evidence base and contextualise findings.
Often literature expansion has been done by navigating the citation network~\cite{Glickman2024AI} however that can miss new papers and reinforces existing citation communities which might be guided by research group bias or narrow areas of research. 

We introduce an agentic system for literature review generation and describe the specific techniques to select and process full-text literature articles to that aim.
We present detailed analysis of how each component contributes to the overall capability of the agentic system using published benchmarks (LitAQ2 and PubMedQA) as well as our own benchmark on literature expansion.

\section{Methodology}
\label{sec:methodology}
Our methodology comprises several key components: document retrieval, information extraction, and verification and diversification. 
Each component is designed to enhance the system's ability to process and analyse scientific literature effectively. For all descriptions in this methodology the LLM used is Anthropic's Claude 3.5 Sonnet~\cite{Claude3}.

\subsection{Document retrieval}
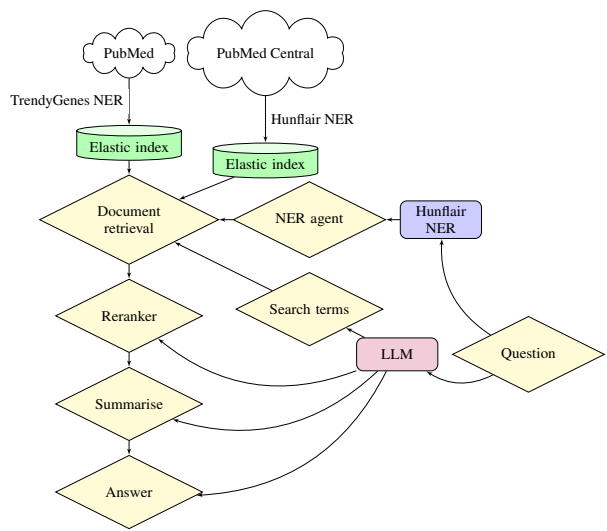
\begin{figure}
\centering
  \resizebox{\columnwidth}{!}{
\begin{tikzpicture}[
    node distance = 0.3cm,
    auto,
    block/.style = {rectangle, draw, fill=blue!20, text width=1.5cm, text centered, rounded corners, minimum height=2em},
    clouds/.style = {cloud, draw, aspect=3},
    database/.style = {cylinder, cylinder uses custom fill, cylinder body fill=green!30, cylinder end fill=green!10, shape border rotate=90, aspect=0.1, draw, text width=2cm, align=center},
    llm/.style = {draw, rectangle, fill=purple!20, text width=1.5cm, align=center, rounded corners, minimum height=2em},
    agent/.style = {draw, diamond, aspect=2.0, fill=yellow!20, text width=2cm, align=center},
    line/.style = {draw, -latex'}
 ]
\node [clouds, font=\small] (s3_1) {PubMed};
\node [clouds, right=of s3_1, font=\small] (s3_2) {PubMed Central};
\node [database, below=1cm of s3_1, font=\small] (db_1) {Elastic index};
\node [database, below=1cm of s3_2, font=\small] (db_2) {Elastic index};
\node[agent, below=of db_1, font=\small](agent_3){Document retrieval};
\node[agent, below=of agent_3, font=\small](agent_4){Reranker};
\node[agent, right=of agent_3, font=\small](agent_1){NER agent};
\node[agent, below=of agent_1, font=\small](agent_2){Search terms};
\node[llm, below right=of agent_2, font=\small](llm){LLM};
\node [block, right=of agent_1, font=\small](hunflair){Hunflair NER};
\node [agent, right=of llm, font=\small](question){Question};
\node[agent, below=of agent_4, font=\small](agent_5){Summarise};
\node[agent, below=of agent_5, font=\small](agent_6){Answer};

\draw [line] (s3_1) -- node[midway, left, fill=white, font=\small] {TrendyGenes NER} (db_1);
\draw [line] (s3_2) -- node[midway, right, fill=white, font=\small] {Hunflair NER} (db_2);
\path [line] (llm) -- (agent_2);
\path [line] (hunflair) -- (agent_1);
\path [line] (agent_1) -- (agent_3);
\path [line] (agent_2) -- (agent_3);
\path [line] (db_1) -- (agent_3);
\path [line] (db_2) -- (agent_3);
\path [line] (agent_3) -- (agent_4) ;
\path [line] (agent_4) -- (agent_5);
\path [line] (agent_5) -- (agent_6);
\path [line] (llm) to [bend left] (agent_4);
\path [line] (llm) to [bend left] (agent_5);
\path [line] (llm) to [bend left] (agent_6);
\path [line] (question) to [bend left] (llm);
\path [line] (question) to [bend left] (hunflair);
\end{tikzpicture}
}
\caption[Data flow from PubMed and PubMed central -> search indices -> document retrieval with LLM -> re-ranking with LLM -> summarisation with LLM -> answering with LLM]{Data flow to support Q\&A}\label{fig:dataflow}
\end{figure}

There are two primary sources of scientific articles used for this evaluation; PubMed~\cite{pubmed} and PubMed Central open access subset~\cite{pmcoa}. PubMed provides citations for $\sim$37 million for biomedical literature (for articles this typically consists of titles and abstract text). The PubMed Central open access contains $\sim$5 million full text biomedical literature entries which is available for machine reading. 

To index the PubMed citations we perform named entity recognition (NER) for diseases and genes and index the titles and abstracts in an Elasticsearch index~\cite{elasticsearch}. 
Genes/gene products often have poor standardisation in naming across literature. Not only can gene names be ambiguous with other entities but also with other genes. 
We use a citation network based approach called \textit{TrendyGenes}~\cite{10.1038/s41598-021-94897-9} that was developed to handle these ambiguities in the NER step.
 
To index the PubMed Central articles we perform NER using \textit{HunFlair}~\cite{10.1093/bioinformatics/btab042,10.48550/arxiv.2402.12372} for genes and diseases.
\textit{Hunflair} provides the classification of spans from which we match the spans to the dictionaries used by \textit{TrendyGenes}~\cite{10.1038/s41598-021-94897-9} in order to normalise to specific entities. 
Each article is indexed as a document with each section in an article handled as sub-document.

In order to retrieve documents to support asking a question we perform the following steps (also see Figure~\ref{fig:dataflow}):
\begin{enumerate}[leftmargin=*]
\item \textbf{\lbrack NER agent\rbrack}\, Run \textit{HunFlair}~\cite{10.1093/bioinformatics/btab042,10.48550/arxiv.2402.12372} NER to identify any gene and disease mentions in the question. 
\item \textbf{\lbrack Search terms agent\rbrack}\, Use the LLM to generate a set of must and should keyword terms to search for matching documents. For should terms we give a boost measure of low, medium and high (see more in Appendix~\ref{sec:search_terms}).
To increase robustness on how the user phrases a question, we generate a set of synonyms for each term which isn't an identifier found by the \textbf{\lbrack NER agent\rbrack}.

\item  \textbf{\lbrack Document retrieval agent\rbrack}\, Using the generated search terms a search template is executed against the Elasticsearch index. This search template also boosts for more recent articles.

\item \textbf{\lbrack Re-ranking agent\rbrack}\, To focus the LLM on more relevant text passages and reduce the amount of text that is included in the prompt, we first chunk each section of each retrieved article into chunks, preserving complete sentences with overlaps between chunks. 
Chunks are sized to $\sim$10000 words with $\sim$250 word overlaps between chunks. 

The LLM is used to generate a set of three diverse proposed answers with rationale and expanded synonyms on keywords (see examples in Appendix~\ref{sec:initial_answers}). 
We then re-rank the article chunks using BM25L~\cite{10.1145/2009916.2010070} with the scores for each proposed answer and the original question we take the combined ranking using the reciprocal rank fusion method~\cite{10.1145/1571941.1572114}. The text is tokenized to lowercase, filtered for stop words and numbers, and stemmed using the Porter2 algorithm. 
This leads to article chunks with relevant information to answering the question being prioritised, which allows to reduce the number of article chunks to be carried forward into the information extraction stage. 
 \end{enumerate}
 
\subsection{Information extraction}
There are three stages applied for the information extraction:
\begin{enumerate}[leftmargin=*]
\item The first stage takes the text chunk retrieved for each article and produces a summary of the key facts related to the question. If there are no relevant facts the article chunk is rejected.
\item The second optional stage creates a prompt which takes subsets of summarised articles (typically in sets of 3) to deduplicate facts across them and then removes any article chunks that provides no new information when compared with the other article chunks in the set.
\item The final stage is to pass the remaining summaries into the LLM context in order to get an answer to the question.
\end{enumerate}

In all these stages the prompts are engineered such that the chain of thought technique is used to improve the model output~\cite{claude:promptengineering}.

\subsection{Verification and diversification}
Chain-of-Verification~\cite{10.48550/arxiv.2309.11495} (CoVe) was originally developed for suppressing incorrect information through taking a draft response, planning and fact checking the draft independently and then giving a final verified response. We extend this framework using the flow in figure~\ref{fig:dataflow} for the draft response and independent fact checking. The intermediate responses are produced in a similar long form format to the original response. The planning stage identifies $\sim$10 key statements in the original response, which are converted into questions to verify their validity. The generated answers, the verification of the original statements and the original response go into generating the final response. We find that the additional questions that are generated in the fact checking stage expand the literature coverage beyond that which was encountered in the original question.

\section{Results}
\subsection{Document retrieval}
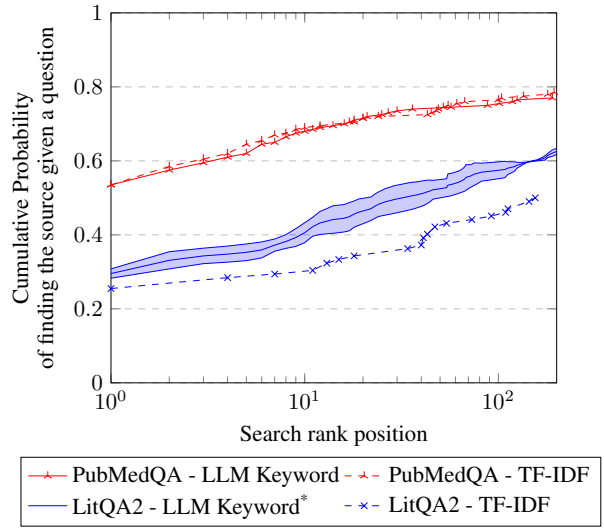
\begin{figure}
\centering
\resizebox{\columnwidth}{!}{
\begin{tikzpicture}
\begin{axis}[
 xmode=log,
 xlabel=Search rank position,
 ylabel style={align=center}, ylabel=Cumulative Probability\\of finding the source given a question,
 xmin=1, xmax=200,
 ymin=0, ymax=1.0,
 legend cell align={left},
 legend style={at={(1.1,-0.2)}}, anchor=north,
 ymajorgrids=true,
 grid style=dashed,
 legend columns=2,
]
\addplot[
color=red,
mark=Mercedes star,
mark options={solid},
] table[x=Rank, y=Cumulative Probability, col sep=comma]{./data/pubmedqa_cdf_source.csv};
\addlegendentry{PubMedQA - LLM Keyword}
\addplot[
color=red,
mark=Mercedes star,
style={dashed},
mark options={solid},
] table[x=Rank, y=Cumulative Probability, col sep=comma]{./data/pubmedqa_bm25_cdf_source.csv};
\addlegendentry{PubMedQA - TF-IDF}
\addplot[
name path=baseline,
color=blue,
style={solid},
] table[x=Rank, y=Mean, col sep=comma]{./data/litqa2_cdf_source.csv};
\addlegendentry{LitQA2 - LLM Keyword\textsuperscript{*}}
\addplot[
name path=baseline_low,
color=blue,
forget plot
] table[x=Rank, y=Low, col sep=comma]{./data/litqa2_cdf_source.csv};
\addplot[
name path=baseline_high,
color=blue,
forget plot
] table[x=Rank, y=High, col sep=comma]{./data/litqa2_cdf_source.csv};
\addplot[blue, opacity=0.2, forget plot] fill between[of=baseline_high and baseline];
\addplot[blue, opacity=0.2, forget plot] fill between[of=baseline_low and baseline];
\addplot[
color=blue,
mark=x,
style={dashed},
mark options={solid},
] table[x=Rank, y=Cumulative Probability, col sep=comma]{./data/litqa2_bm25_cdf_source.csv};
\addlegendentry{LitQA2 - TF-IDF}
\end{axis}
\end{tikzpicture}
}
\caption{Search rank with search term agent for benchmark sources in PubMed and PubMed Central (* with 95\% CI)}\label{fig:cdf_rank}
\end{figure}

We evaluate the document retrieval by determining where in the search ranking the source documents for the benchmarks appear. 
For evaluating the retrieval on full text articles in the PubMed Central Open Access dataset we can only consider a subset of the LitQA2 benchmark~\cite{10.48550/arxiv.2407.10362} (for which only 103 out of 199 are found in the PubMed Central OA subset). 
Similarly for evaluating retrieval on PubMed we take the first 200 labelled questions from the PubMedQA benchmark~\cite{10.48550/arxiv.1909.06146} to see where their sources appear in the search ranking.
We only consider up to the 200\textsuperscript{th} ranked search result for both sources in this evaluation. 
We compare our approach to a baseline using a simpler `More Like This' query, which leverages TF-IDF to measure document similarity. For LitQA2, we include answer options in the query. In the PubMedQA case we only evaluated a small subset of the test set as it became evident there was little gain in using the LLM keyword approach versus the baseline approach. This is not very surprising given the questions are directly generated from the titles of the articles~\cite{10.48550/arxiv.1909.06146}. Figure~\ref{fig:cdf_rank} shows that for the LLM generated keyword approach LitQA2 benchmark 63.6$\pm$1.2 source articles were found in the top 200 search hits out of the possible 103 compared with 52 for the simpler approach. For PubMedQA only 154 source articles were found out of the 200 questions tested. The top 200 was chosen as a cutoff for performance reasons, to be able to do the re-ranking in memory.

\subsection{Re-ranking}
\begin{figure}
\centering
\resizebox{\columnwidth}{!}{
\begin{tikzpicture}
\begin{axis}[
xmode=log,
 xlabel=Chunk rank position,
 ylabel style={align=center}, ylabel=Cumulative Probability\\of finding the key passage across\\ all chunks,
 ymin=0, ymax=1.0,
 xmin=1, xmax=1000,
 legend pos=south east, legend cell align={left},
 ymajorgrids=true,
 grid style=dashed,
  unbounded coords=jump,
]
\addplot[
name path=reranked,
color=black,
] table[x=Rank, y=Mean, col sep=comma]{./data/cdf_rerank_reranked.csv};
\addlegendentry{Reranked}
\addplot[
name path=reranked_low,
color=black,
forget plot
] table[x=Rank, y=Low, col sep=comma]{./data/cdf_rerank_reranked.csv};
\addplot[
name path=reranked_high,
color=black,
forget plot
] table[x=Rank, y=High, col sep=comma]{./data/cdf_rerank_reranked.csv};
\addplot[black, opacity=0.2, forget plot] fill between[of=reranked_high and reranked];
\addplot[black, opacity=0.2, forget plot] fill between[of=reranked_low and reranked];
\addplot[
name path=baseline,
color=blue,
style={dashed},
] table[x=Rank, y=Mean, col sep=comma]{./data/cdf_rerank_baseline.csv};
\addlegendentry{Baseline}
\addplot[
name path=baseline_low,
color=blue,
forget plot
] table[x=Rank, y=Low, col sep=comma]{./data/cdf_rerank_baseline.csv};
\addplot[
name path=baseline_high,
color=blue,
forget plot
] table[x=Rank, y=High, col sep=comma]{./data/cdf_rerank_baseline.csv};
\addplot[blue, opacity=0.2, forget plot] fill between[of=baseline_high and baseline];
\addplot[blue, opacity=0.2, forget plot] fill between[of=baseline_low and baseline];
\end{axis}
\end{tikzpicture}
}
\caption{LitQA2 key passage location with re-ranking (with 95\% CI)}\label{fig:reranking}
\end{figure}
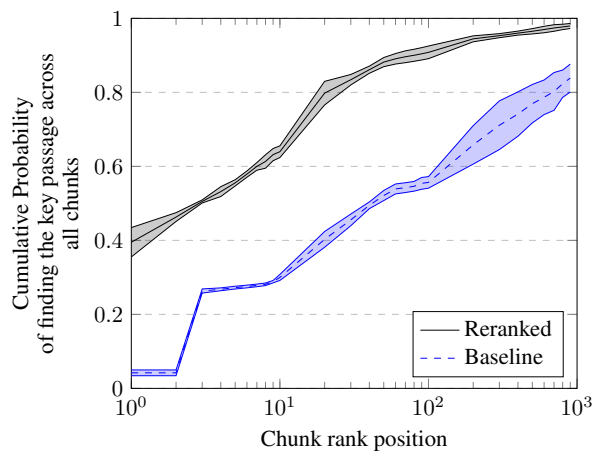

For the re-ranking we create a baseline by taking every section in all 200 full-text articles for a given question, as found in figure~\ref{fig:cdf_rank}, and then chunk each section as described in ~\ref{sec:methodology}. 
In the LitQA2 benchmark~\cite{10.48550/arxiv.2407.10362} there is a key passage of text assigned to each question which is required to answer it. We find the position in the ranking of chunks at which the key passage appears. 
We contrast this with re-ranking each section across all publications in figure~\ref{fig:reranking}. It can be seen that within the first 30 chunks $\sim$83\% of the key passages can be found with the re-ranking, this is in contrast to $\sim$900 chunks for the baseline. 

\subsection{Information extraction}
For the LitQA2 benchmark~\cite{10.48550/arxiv.2407.10362} we consider the split between question sources which are available/not available for machine reading within the PubMed Central OA dataset~\cite{pmcoa}. In the assessment of the benchmark we follow the same conventions as in the original paper~\cite{10.48550/arxiv.2407.10362}:
\begin{itemize}[leftmargin=*]
\item Accuracy = Total correct answers / Total questions
\item Coverage = Total attempted answers / Total questions
\item Precision = Total correct answers / Total attempted answers
\end{itemize}

To get a better understanding of the LLM's comprehension and potential information loss we evaluate how good the model is at answering the question given the source, by just feeding the original article source (table~\ref{tab:LitQA2_source_evaluations}). 
We provide in the LLM context different levels of granularity: the key passage given by the benchmark, the section of the article which contains the key passage, the full article and a context specific summarisation of the full article. 
According to the LitQA2 benchmark paper the precision on these all exceed human levels, a finding which we recapitulate here.

\begin{table}
  \centering
    \resizebox{\columnwidth}{!}{
    \begin{tabular}{lccc}
  \hline
  Step & Recall of source article (\% $\pm$ 95\% CI) \\
   \hline
   LLM Keyword retrieval (top 200) & 61.7 $\pm$ 1.2 \\
   Re-ranking (top 30 chunks) & 57.3 $\pm$ 1.9 \\
   LLM context summarisation and retention & 57.0 $\pm$ 2.4 \\
   Attribution to answer & 50.5 $\pm$ 0.8 \\
   \hline
   \end{tabular}
   }
\caption{LitQA2 source article recall}\label{tab:LitQA2_efficiency}
\end{table}
  
\subsection{Question answering}
The LitQA2 benchmark~\cite{10.48550/arxiv.2407.10362} assumes that only the source paper contains the information to achieve the correct answer. We find that this assumption does not hold over our text corpus.

The recall on the source paper for the open access subset is provided in Table~\ref{tab:LitQA2_efficiency}, from which we can conclude that the highest loss in recall occurs in the initial paper retrieval stage. The LLM context summarisation and retention step allows for the LLM to choose to drop what it considers are irrelevant papers. The final answering stage allows the LLM to attribute the answer to a single paper and this is what leads to our 50.5\% final recall. In the table the mean and 95\% confidence intervals of the results of 3 repeats are provided. In our evaluated subset, $\sim$10 answers are correct with different paper attributions, suggesting that the assumption of specificity for the benchmark question might not be correct. 

Table~\ref{tab:LitQA2_source_evaluations} provides the evaluation for our agentic literature review system on the LitQA2 benchmark.  The assumption of a single source paper containing the answer to a benchmark question equates to a maximum potential accuracy of $\sim$50\% based on the open subset of the benchmark, assuming that the top 30 re-ranked chunks all equate to correct answers.
  
The coverage is higher than that attributable to the source papers, however additional papers which allowed the LLM to answer the question increased the coverage in total 60.4$\pm$0.6\% questions with other papers used to answer the questions. 

\begin{table}
\centering
    \resizebox{\columnwidth}{!}{
\begin{tabular}{lccc}
  \hline
  Information provided in LLM context & Accuracy & Coverage & Precision \\
   \hline
Key passage & 0.955 & 0.990 & 0.965 \\
Key section (available) & 0.983 & 1.000 & 0.982 \\
All sections (available) & 0.990 & 1.000 & 0.990 \\
Summarised article (available) & 0.970 & 1.000 & 0.970 \\
Our full agentic workflow (available)\textsuperscript{*} & 0.604$\pm$0.006 & 0.668$\pm$0.007 & 0.903$\pm$0.019 \\
PaperQA2 (all) & 0.66 & 0.78 & 0.852 \\
\hline
\end{tabular}
}
 \caption{Evaluation of information loss at different stages and comparison with LitQA2 benchmark (* with 95\% CI)}
 \label{tab:LitQA2_source_evaluations}
 \end{table}

\subsection{Verification and diversification}
  In order to benchmark this capability we consider 10 recent review articles published in Nature and the references used with in them. 
  For each review we get an LLM to generate the key question being asked from the abstract of the paper and use that as our starting point (see more in Appendix~\ref{sec:nature_review_dataset}). 
  We perform the retrieval with the date cutoffs set as the publication date of the reviews. 
  The benchmark is then measured as the recall on the references on each paper. We consider the references which either come out of the draft response generation or in the fact checking retrieval process.
  We consider both the direct references used in the reviews and the secondary references (from which the primary research might come). 
  Table~\ref{tab:COV_recall} shows that the mean coverage of primary and secondary references on the Nature review articles improves with 6\% to 25\% with the CoVe approach. 
  
 \begin{table}
  \centering
    \resizebox{\columnwidth}{!}{
  \begin{filecontents*}{./data/cov_literature_coverage_nature_reviews.csv}
PMID,Initial recall,Initial with parents recall,All recall,All with parents recall,Total references
38570716,1.6260162601626018,12.195121951219512,13.821138211382115,56.09756097560976,123
37758888,2.9629629629629632,33.33333333333333,14.814814814814813,48.148148148148145,135
38418916,1.1235955056179776,9.550561797752808,5.056179775280898,39.325842696629216,178
37138108,1.0526315789473684,2.1052631578947367,14.736842105263156,33.68421052631579,95
37344646,0.0,5.641025641025641,2.564102564102564,25.64102564102564,195
37100941,0.0,3.1746031746031744,5.555555555555555,19.047619047619047,126
38326590,0.0,0.7246376811594203,1.4492753623188406,14.492753623188406,138
38750233,0.0,0.6097560975609756,0.0,5.487804878048781,164
37316719,0.0,0.0,0.5586592178770949,3.910614525139665,179
37612393,0.7633587786259541,0.7633587786259541,1.5267175572519083,3.0534351145038165,131

\end{filecontents*}
\pgfplotstabletypeset[
          col sep=comma,
          ignore chars={"},
          fixed, precision=0,
          fixed zerofill,
          every head row/.style={before row=\toprule,after row=\midrule},
          every last row/.style={after row=\bottomrule},
          columns/PMID/.style={column type={l}, string type},
          columns/Initial recall/.style={column type={c}, column name={\thead{Initial\\recall (\%)}}},
          columns/Initial with parents recall/.style={column type={c}, column name={\thead{Initial recall\\ with secondary (\%)}}},
	  columns/All recall/.style={column type={c}, column name={\thead{CoVe\\ recall (\%)}}},
          columns/All with parents recall/.style={column type={c}, column name={\thead{CoVe recall\\ with secondary (\%)}}},
          columns={PMID,Initial recall,All recall,Initial with parents recall,All with parents recall},
  ]{./data/cov_literature_coverage_nature_reviews.csv}
  }
\caption{Recall of references with CoVe} \label{tab:COV_recall}
\end{table}

\section{Discussion}
We have demonstrated that when compared with the PaperQA2 paper~\cite{paperqa2} we achieved a slightly higher attribution to original source for correct answers of 50.4\% compared with 48.2\%. When we consider the accuracy of answering the questions overall we achieved 60.4\% compared with 66\%. However, comparing for precision we exceed PaperQA2 achieving 90\% compared with 85\%.
  
In the LAB-Bench~\cite{10.48550/arxiv.2407.10362} paper it can be seen that in answering the benchmark Claude 3.5 Sonnet is significantly more cautious at choosing whether to give an answer compared to GPT-4o (0.12 vs 0.58 coverage of the LitQA2 benchmark). We hypothesise this baseline caution leads to our observed reduced coverage compared with PaperQA2, which is built on GPT-4o, this in part leads to our reduced accuracy overall, however, our precision comes out slightly higher. It should be possible to force the model to take a guess to drive up coverage, but as the retrieval and attribution of our system is similar to that of LitQA2, this would mean the model would make a guess that isn't based on having appropriate information in the context. Given that there is a 25\% chance of picking the correct answer in a multiple-choice test, we would be gaming the system rather than testing the system's capability of answering questions based on information from the literature. Therefore, the key area for improvement on the benchmark lies in the initial retrieval stage, both for our system as well as PaperQA2 (We make further analysis of this in appendix~\ref{sec:further_benchmark_analysis}).
 
The LitQA2 benchmark was designed for answering very specific questions with multiple-choice answers, whereas our literature review system was designed to summarise and contextualise a broad literature base around a given research question.
Generating meaningful benchmarks is difficult, as was also shown in our results where the benchmark assumption of a question only being answerable from a single paper was broken several times. This also leads to the concern that if more than one paper can provide a relevant answer, there could be papers with contradictory claims that affect the answer. Our system has been designed to include a verification and diversification step to handle such situation, by including a wider range of relevant papers for question answering. Furthermore with the query expansion and synonym handling (particularly of genes/proteins) provided in our methods these are not particularly demonstrated well in the benchmark. The benchmark questions rarely deviate from the keywords used in the passages to generate them such that the retrieval systems are not particularly pushed hard to resolve semantically similar terms.
Despite benchmarks necessarily being limited compared to real-world applications, they serve a useful purpose for enabling comparisons and improving methods.

While our methodology does not exceed the accuracy measurement of PaperQA2~\cite{paperqa2} it does demonstrate a technique which gets similar levels of attribution and retrieval without the need to exploit the citation network, and without the use of dense retrieval methods. 
This makes the technique transferable to document stores where this information is not available. 

For current agentic systems producing a well-referenced review article there is a strong trade-off between execution time and the quantity of literature processed in such systems. 
The strength of LLMs to extract information from papers as displayed in table~\ref{tab:LitQA2_source_evaluations} is already very good and for a single paper faster than a typical human. 
Being able to benchmark the quality of a review for an area of literature is difficult as the generated text is often far larger than you might generate for bench-marking summarisation tasks and the quality is also far more based on human opinion. 
Furthermore, references used in a review article will be biased to the authors knowledge and affiliations which an agentic system like ours will not be able to reproduce. 
Even so we were able to demonstrate significant improvement to the literature coverage by use of the CoVe technique. In reality the end-to-end time to perform such agentic literature reviews typically takes $\sim$10-30 minutes. 
This is significantly slower than many of the commercial chat apps (such as ChatGPT) however the quality and grounding of responses (as qualitatively accessed by users) is often a lot better. 
When compared with asking a human to perform a literature review our agentic literature review system is significantly faster.

\section{Conclusion}
We developed a methodology for answering questions from scientific literature and validated the retrieval and answering capabilities of these methods against the LitQA2 benchmark. 
We have further demonstrated a method for exploring a wider literature space when generating scientific literature reviews using CoVe.
We validated this approach by assessing the coverage of relevant literature. 
Scaling these methods allows for scientists to explore more literature in a shorter amount of time than would be possible previously.

\bibliographystyle{ieeetr}
\bibliography{references}

\begin{thebibliography}{10}

\bibitem{10.48550/arxiv.2108.07258}
R.~Bommasani, D.~A. Hudson, E.~Adeli, R.~Altman, S.~Arora, S.~v. Arx, M.~S. Bernstein, J.~Bohg, A.~Bosselut, E.~Brunskill, E.~Brynjolfsson, S.~Buch, D.~Card, R.~Castellon, N.~Chatterji, A.~Chen, K.~Creel, J.~Q. Davis, D.~Demszky, C.~Donahue, M.~Doumbouya, E.~Durmus, S.~Ermon, J.~Etchemendy, K.~Ethayarajh, L.~Fei-Fei, C.~Finn, T.~Gale, L.~Gillespie, K.~Goel, N.~Goodman, S.~Grossman, N.~Guha, T.~Hashimoto, P.~Henderson, J.~Hewitt, D.~E. Ho, J.~Hong, K.~Hsu, J.~Huang, T.~Icard, S.~Jain, D.~Jurafsky, P.~Kalluri, S.~Karamcheti, G.~Keeling, F.~Khani, O.~Khattab, P.~W. Koh, M.~Krass, R.~Krishna, R.~Kuditipudi, A.~Kumar, F.~Ladhak, M.~Lee, T.~Lee, J.~Leskovec, I.~Levent, X.~L. Li, X.~Li, T.~Ma, A.~Malik, C.~D. Manning, S.~Mirchandani, E.~Mitchell, Z.~Munyikwa, S.~Nair, A.~Narayan, D.~Narayanan, B.~Newman, A.~Nie, J.~C. Niebles, H.~Nilforoshan, J.~Nyarko, G.~Ogut, L.~Orr, I.~Papadimitriou, J.~S. Park, C.~Piech, E.~Portelance, C.~Potts, A.~Raghunathan, R.~Reich, H.~Ren, F.~Rong, Y.~Roohani, C.~Ruiz, J.~Ryan, C.~R{\'e},
  D.~Sadigh, S.~Sagawa, K.~Santhanam, A.~Shih, K.~Srinivasan, A.~Tamkin, R.~Taori, A.~W. Thomas, F.~Tram{\`e}r, R.~E. Wang, W.~Wang, B.~Wu, J.~Wu, Y.~Wu, S.~M. Xie, M.~Yasunaga, J.~You, M.~Zaharia, M.~Zhang, T.~Zhang, X.~Zhang, Y.~Zhang, L.~Zheng, K.~Zhou, and P.~Liang, ``{On the Opportunities and Risks of Foundation Models},'' {\em arXiv}, 2021.

\bibitem{10.48550/arxiv.2303.12712}
S.~Bubeck, V.~Chandrasekaran, R.~Eldan, J.~Gehrke, E.~Horvitz, E.~Kamar, P.~Lee, Y.~T. Lee, Y.~Li, S.~Lundberg, H.~Nori, H.~Palangi, M.~T. Ribeiro, and Y.~Zhang, ``{Sparks of Artificial General Intelligence: Early experiments with GPT-4},'' {\em arXiv}, 2023.

\bibitem{10.1038/s41586-023-06792-0}
D.~A. Boiko, R.~MacKnight, B.~Kline, and G.~Gomes, ``{Autonomous chemical research with large language models},'' {\em Nature}, vol.~624, no.~7992, pp.~570--578, 2023.

\bibitem{bran2023chemcrowaugmentinglargelanguagemodels}
A.~M. Bran, S.~Cox, O.~Schilter, C.~Baldassari, A.~D. White, and P.~Schwaller, ``Chemcrow: Augmenting large-language models with chemistry tools,'' 2023.

\bibitem{lala2023paperqa}
J.~L{\'a}la, O.~O'Donoghue, A.~Shtedritski, S.~Cox, S.~G. Rodriques, and A.~D. White, ``Paperqa: Retrieval-augmented generative agent for scientific research,'' {\em arXiv preprint arXiv:2312.07559}, 2023.

\bibitem{10.48550/arxiv.2408.01107}
C.~Wang, Q.~Long, M.~Xiao, X.~Cai, C.~Wu, Z.~Meng, X.~Wang, and Y.~Zhou, ``{BioRAG: A RAG-LLM Framework for Biological Question Reasoning},'' {\em arXiv}, 2024.

\bibitem{NEURIPS2020_6b493230}
P.~Lewis, E.~Perez, A.~Piktus, F.~Petroni, V.~Karpukhin, N.~Goyal, H.~K\"{u}ttler, M.~Lewis, W.-t. Yih, T.~Rockt\"{a}schel, S.~Riedel, and D.~Kiela, ``Retrieval-augmented generation for knowledge-intensive nlp tasks,'' in {\em Advances in Neural Information Processing Systems} (H.~Larochelle, M.~Ranzato, R.~Hadsell, M.~Balcan, and H.~Lin, eds.), vol.~33, pp.~9459--9474, Curran Associates, Inc., 2020.

\bibitem{gao2024retrievalaugmentedgenerationlargelanguage}
Y.~Gao, Y.~Xiong, X.~Gao, K.~Jia, J.~Pan, Y.~Bi, Y.~Dai, J.~Sun, M.~Wang, and H.~Wang, ``Retrieval-augmented generation for large language models: A survey,'' 2024.

\bibitem{10.48550/arxiv.2312.07559}
J.~Lála, O.~O'Donoghue, A.~Shtedritski, S.~Cox, S.~G. Rodriques, and A.~D. White, ``{PaperQA: Retrieval-Augmented Generative Agent for Scientific Research},'' {\em arXiv}, 2023.

\bibitem{paperqa2}
M.~D. Skarlinski, S.~Cox, J.~M. Laurent, J.~D. Braza, M.~Hinks, M.~J. Hammerling, M.~Ponnapati, S.~G. Rodriques, and A.~D. White, ``{LANGUAGE AGENTS ACHIEVE SUPERHUMAN SYNTHESIS OF SCIENTIFIC KNOWLEDGE},'' 9 2024.

\bibitem{10.48550/arxiv.1909.06146}
Q.~Jin, B.~Dhingra, Z.~Liu, W.~W. Cohen, and X.~Lu, ``{PubMedQA: A Dataset for Biomedical Research Question Answering},'' {\em arXiv}, 2019.

\bibitem{10.48550/arxiv.2407.10362}
J.~M. Laurent, J.~D. Janizek, M.~Ruzo, M.~M. Hinks, M.~J. Hammerling, S.~Narayanan, M.~Ponnapati, A.~D. White, and S.~G. Rodriques, ``{LAB-Bench: Measuring Capabilities of Language Models for Biology Research},'' {\em arXiv}, 2024.

\bibitem{10.48550/arxiv.2403.17752}
W.~Li, L.~Li, T.~Xiang, X.~Liu, W.~Deng, and N.~Garcia, ``{Can multiple-choice questions really be useful in detecting the abilities of LLMs?},'' {\em arXiv}, 2024.

\bibitem{10.48550/arxiv.2406.07545}
A.~Myrzakhan, S.~M. Bsharat, and Z.~Shen, ``{Open-LLM-Leaderboard: From Multi-choice to Open-style Questions for LLMs Evaluation, Benchmark, and Arena},'' {\em arXiv}, 2024.

\bibitem{Glickman2024AI}
M.~Glickman and Y.~Zhang, ``{AI} and {Generative} {AI} for {Research} {Discovery} and {Summarization},'' {\em Harvard Data Science Review}, vol.~6, apr 30 2024.
\newblock https://hdsr.mitpress.mit.edu/pub/xedo5giw.

\bibitem{neelakantan2022textcodeembeddingscontrastive}
A.~Neelakantan, T.~Xu, R.~Puri, A.~Radford, J.~M. Han, J.~Tworek, Q.~Yuan, N.~Tezak, J.~W. Kim, C.~Hallacy, J.~Heidecke, P.~Shyam, B.~Power, T.~E. Nekoul, G.~Sastry, G.~Krueger, D.~Schnurr, F.~P. Such, K.~Hsu, M.~Thompson, T.~Khan, T.~Sherbakov, J.~Jang, P.~Welinder, and L.~Weng, ``Text and code embeddings by contrastive pre-training,'' 2022.

\bibitem{douze2024faisslibrary}
M.~Douze, A.~Guzhva, C.~Deng, J.~Johnson, G.~Szilvasy, P.-E. Mazaré, M.~Lomeli, L.~Hosseini, and H.~Jégou, ``The faiss library,'' 2024.

\bibitem{10.1561/1500000019}
S.~Robertson and H.~Zaragoza, ``The probabilistic relevance framework: Bm25 and beyond,'' {\em Found. Trends Inf. Retr.}, vol.~3, p.~333–389, Apr. 2009.

\bibitem{formal2021spladesparselexicalexpansion}
T.~Formal, B.~Piwowarski, and S.~Clinchant, ``Splade: Sparse lexical and expansion model for first stage ranking,'' 2021.

\bibitem{https://doi.org/10.1155/2014/132158}
A.~R. Rivas, E.~L. Iglesias, and L.~Borrajo, ``Study of query expansion techniques and their application in the biomedical information retrieval,'' {\em The Scientific World Journal}, vol.~2014, no.~1, p.~132158, 2014.

\bibitem{jiang2024longragenhancingretrievalaugmentedgeneration}
Z.~Jiang, X.~Ma, and W.~Chen, ``Longrag: Enhancing retrieval-augmented generation with long-context llms,'' 2024.

\bibitem{Claude3}
Anthropic, ``The claude 3 model family: Opus, sonnet, haiku,'' 2024.

\bibitem{liu-etal-2024-lost}
N.~F. Liu, K.~Lin, J.~Hewitt, A.~Paranjape, M.~Bevilacqua, F.~Petroni, and P.~Liang, ``Lost in the middle: How language models use long contexts,'' {\em Transactions of the Association for Computational Linguistics}, vol.~12, pp.~157--173, 2024.

\bibitem{needleinahaystack}
G.~Kamradt, ``I pressure tested gpt-4's 128k context retrieval.'' https://www.youtube.com/watch?v=KwRRuiCCdmc, November 2023.

\bibitem{ANNAMORADNEJAD2024123685}
I.~Annamoradnejad and G.~Zoghi, ``Colbert: Using bert sentence embedding in parallel neural networks for computational humor,'' {\em Expert Systems with Applications}, vol.~249, p.~123685, 2024.

\bibitem{patel2024retrieveexplainevidencedrivenpredictions}
R.~Patel, A.~Brayne, R.~Hintzen, D.~Jaroslawicz, G.~Neculae, and D.~Corneil, ``Retrieve to explain: Evidence-driven predictions with language models,'' 2024.

\bibitem{10.1145/2009916.2010070}
Y.~Lv and C.~Zhai, ``When documents are very long, bm25 fails!,'' in {\em Proceedings of the 34th International ACM SIGIR Conference on Research and Development in Information Retrieval}, SIGIR '11, (New York, NY, USA), p.~1103–1104, Association for Computing Machinery, 2011.

\bibitem{pubmed}
{Bethesda (MD): National Library of Medicine}, ``{PubMed}.'' internet, August 2024.

\bibitem{pmcoa}
{Bethesda (MD): National Library of Medicine}, ``{PMC Open Access Subset}.'' internet, August 2024.

\bibitem{elasticsearch}
{Elastic NV}, ``Elasticsearch.'' \url{https://www.elastic.co/elasticsearch/}, 2024.
\newblock Accessed: [2024/08/20].

\bibitem{10.1038/s41598-021-94897-9}
G.~S. N{\'a}jera, D.~N. Carl{\'o}n, and D.~J. Crowther, ``{TrendyGenes, a computational pipeline for the detection of literature trends in academia and drug discovery},'' {\em Scientific Reports}, vol.~11, no.~1, p.~15747, 2021.

\bibitem{10.1093/bioinformatics/btab042}
L.~Weber, M.~S{\"a}nger, J.~M{\"u}nchmeyer, M.~Habibi, U.~Leser, and A.~Akbik, ``{HunFlair: an easy-to-use tool for state-of-the-art biomedical named entity recognition},'' {\em Bioinformatics}, vol.~37, no.~17, pp.~2792--2794, 2021.

\bibitem{10.48550/arxiv.2402.12372}
M.~S{\"a}nger, S.~Garda, X.~D. Wang, L.~Weber-Genzel, P.~Droop, B.~Fuchs, A.~Akbik, and U.~Leser, ``{HunFlair2 in a cross-corpus evaluation of biomedical named entity recognition and normalization tools},'' {\em arXiv}, 2024.

\bibitem{10.1145/1571941.1572114}
G.~V. Cormack, C.~L.~A. Clarke, and S.~Buettcher, ``Reciprocal rank fusion outperforms condorcet and individual rank learning methods,'' in {\em Proceedings of the 32nd International ACM SIGIR Conference on Research and Development in Information Retrieval}, SIGIR '09, (New York, NY, USA), p.~758–759, Association for Computing Machinery, 2009.

\bibitem{claude:promptengineering}
Anthropic, ``Prompt engineering guidance for claude.'' https://docs.anthropic.com/en/docs/build-with-claude/prompt-engineering/overview, August 2024.

\bibitem{10.48550/arxiv.2309.11495}
S.~Dhuliawala, M.~Komeili, J.~Xu, R.~Raileanu, X.~Li, A.~Celikyilmaz, and J.~Weston, ``{Chain-of-Verification Reduces Hallucination in Large Language Models},'' {\em arXiv}, 2023.

\end{thebibliography}

\appendix

\section{Further benchmark analysis}
\label{sec:further_benchmark_analysis}

\begin{figure}[htbp]
\centering
\resizebox{\columnwidth}{!}{
\begin{tikzpicture}
\begin{axis}[
 xmode=log,
 xlabel=Search rank position,
 ylabel style={align=center}, ylabel=Cumulative Probability\\of finding the source given a question,
 xmin=1, xmax=4000,
 ymin=0, ymax=1.0,
 legend cell align={left},
 anchor=north,
 ymajorgrids=true,
 grid style=dashed,
 legend columns=2,
]
\addplot[
name path=baseline,
color=blue,
style={solid},
] table[x=Rank, y=Mean, col sep=comma]{./data/litqa2_exhaustive_retrieval.csv};
\addlegendentry{LitQA2 - LLM Keyword\textsuperscript{*}}
\addplot[
name path=baseline_low,
color=blue,
forget plot
] table[x=Rank, y=Low, col sep=comma]{./data/litqa2_exhaustive_retrieval.csv};
\addplot[
name path=baseline_high,
color=blue,
forget plot
] table[x=Rank, y=High, col sep=comma]{./data/litqa2_exhaustive_retrieval.csv};
\addplot[blue, opacity=0.2, forget plot] fill between[of=baseline_high and baseline];
\addplot[blue, opacity=0.2, forget plot] fill between[of=baseline_low and baseline];
\end{axis}
\end{tikzpicture}
}
\caption{Search rank with search term agent for benchmark sources in PubMed Central (* with 95\% CI)}\label{fig:cdf_rank_extended}
\end{figure}
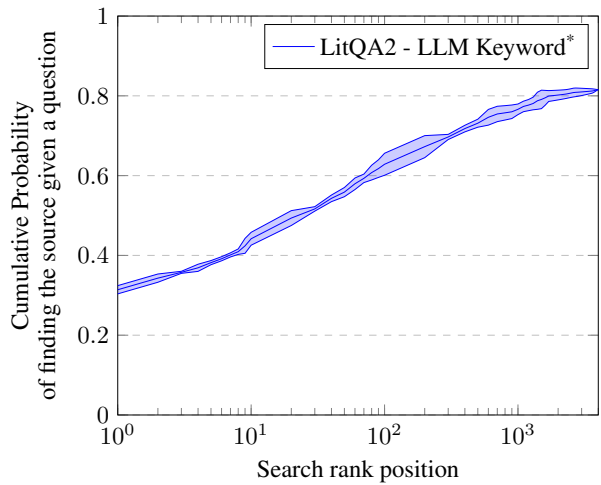

In both PaperQA2 and our methods we demonstrate similar initial retrieval. In figure~\ref{fig:cdf_rank_extended} we explore the search rank up to the first 4000 entries. It can be seen that beyond the first 100 entries there is a very long tail. 

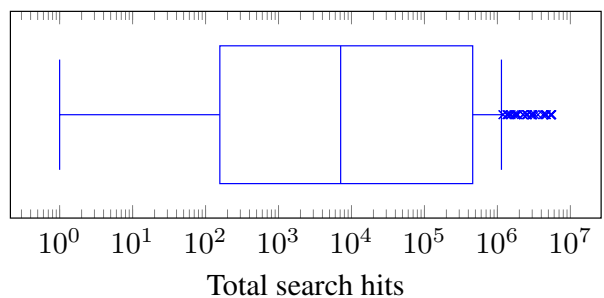
\begin{figure}[htbp]
\centering
\resizebox{\columnwidth}{!}{
\begin{tikzpicture}
\begin{axis}[
    boxplot,
    y=2cm,
    ytick=\empty,
    table/y=Total,
    xlabel=Total search hits,
    boxplot/draw direction=x,
    xmode=log,
    mark=x,
    ymin=0.4, ymax=1.6
]
 \addplot[blue] table{./data/totals_exhaustive_retrieval.csv};
\end{axis}
\end{tikzpicture}
}
\caption{Total potential search hits across questions in LitQA2 with search term agent}\label{fig:total_matches}
\end{figure}

In figure~\ref{fig:total_matches} the aggregated distribution of total search hits based on the generated keyword search are shown for 3 repeats of the LitQA2 benchmark (considering only the available papers). It can be seen that the distribution has a long tail going out into millions of hits. Given both PaperQA2 (which relies on Semantic Scholar's sparse retrieval with a dense reranker) and our sparse retrieval techniques exhibit similar initial retrieval it would suggest that the retrieval issues don't reside solely in the methods alone but potentially the specificity of the questions themselves.

\section{Verification and diversification analysis}
\label{sec:nature_review_dataset}
For the analysis of the CoVe we use a set of recent Nature reviews to formulate a set of starting questions. The derived starting questions can be seen in table~\ref{tab:nature_review_dataset}.

In the CoVe planning step we get the LLM to propose a series of follow up questions based on the initial answer. For the question \textit{How can the integration of cryo-EM and cryo-electron tomography bridge the gap between structural and cell biology to provide a comprehensive understanding of macromolecular interactions within their native cellular context?} a subset of the resulting questions are:
{\small
\begin{itemize}[leftmargin=*]
    \item What is the highest resolution achieved by single-particle cryo-EM for protein structures?
    \item Can cryo-ET provide high-resolution structural details comparable to single-particle cryo-EM?
    \item What specific cellular interactions or structures can be observed using integrated cryo-EM and cryo-ET that cannot be seen with traditional structural biology methods?
    \item Are there any limitations or potential artifacts in fitting high-resolution cryo-EM structures into lower-resolution cryo-ET maps?
    \item Can you provide specific examples of transient cellular processes or interactions that have been uniquely revealed by cryo-EM or cryo-ET?
    \item What is the typical thickness of lamellae prepared by FIB milling for cryo-ET, and how does this compare to the maximum sample thickness for direct cryo-ET imaging?
    \item What is the highest resolution achieved to date using subtomogram averaging in cryo-ET, and how does it compare to single-particle cryo-EM?
    \item How does the structural information obtained from the cryoWriter system compare to that from traditional cryo-EM of purified complexes in terms of resolution and native state preservation?
\end{itemize}
}
These questions are nuanced as they are based on information retrieved from the initial round of retrieval and summarisation. 

\section{Search term query expansion}
\label{sec:search_terms}
In the search term query expansion we ask the LLM to map to known identifiers found through NER and LLM identified keywords, for which it generates synonyms. 

For example, for the question \textit{Relative to wild type zebrafish embryos, those with Kif9 knocked down display what ciliary phenotype?} the following terms are generated:
\begin{lstlisting}[basicstyle=\tiny,language=xml,tabsize=2,stringstyle=\color{red},emph={must,should},emphstyle={\color{magenta}},breaklines=true,frame=single]
<terms>
  <must>
    <term>zebrafish</term>
    <term>Danio rerio</term>
  </must>
  <must>
    <term>cilia</term>
    <term>ciliary</term>
  </must>
  <must>
    <term>ENSEMBL:ENSG00000088727</term>
  </must>
  <should>
    <term>embryo</term>
    <term>embryonic</term>
    <boost>MEDIUM</boost>
  </should>
  <should>
    <term>knockdown</term>
    <term>knocked down</term>
    <term>silencing</term>
    <term>gene silencing</term>
    <boost>HIGH</boost>
  </should>
  <should>
    <term>phenotype</term>
    <term>morphology</term>
    <boost>LOW</boost>
  </should>
  <should>
    <term>wild type</term>
    <term>WT</term>
    <boost>MEDIUM</boost>
  </should>
</terms>
\end{lstlisting}
where each <should> and <must> are the terms with their synonyms or the term matched by the NER. The XML structure uses <must> tags for required terms and <should> tags for optional terms that may improve search results. The <boost> values indicate a weighting to apply to should terms in the search scoring.

\section{Initial answers for reranking}
\label{sec:initial_answers}
In order to perform re-ranking in addition to the question we use 3 plausible answers making use of the LLMs own memory to get sample text in a form which is similar to what a long form answer might look like (even if not correct). These are then used to query the BM25L index. The outputs also include synonyms of key terms to try to overcome some of the semantic short comings typically associated to sparse retrieval.

For example, for the question \textit{Relative to wild type zebrafish embryos, those with Kif9 knocked down display what ciliary phenotype?} the following answers are given:
\begin{itemize}[label={},leftmargin=*]
\item {\small Kif9 knockdown zebrafish embryos likely display shorter cilia (synonyms/abbreviations: reduced-length cilia, truncated cilia, diminished cilia, abbreviated cilia, stunted cilia) as the ciliary phenotype. Kif9 is a kinesin motor protein (synonyms/abbreviations: molecular motor, motor enzyme, KIF, kinesin-like protein, microtubule-associated motor) that plays a role in intraflagellar transport [synonyms/abbreviations: IFT, ciliary transport, flagellar transport, cilia protein trafficking, axonemal transport]. Without Kif9, the transport of building materials to the ciliary tip may be impaired, resulting in shorter cilia due to insufficient elongation during ciliogenesis (synonyms/abbreviations: cilia formation, ciliary assembly, cilia biogenesis, cilia growth, flagella formation).}
\item {\small The ciliary phenotype observed in Kif9 knockdown zebrafish embryos is likely increased beat frequency (synonyms/abbreviations: accelerated ciliary motion, enhanced ciliary beating, rapid cilia oscillation, heightened ciliary activity, faster cilia movement). Kif9 may function as a regulatory protein [synonyms/abbreviations: modulator, control factor, regulator molecule, governing protein, homeostatic agent] that normally dampens ciliary beating. In its absence, the cilia might lack proper regulation, leading to increased beat frequency. This could result in altered fluid flow (synonyms/abbreviations: hydrodynamic changes, fluid dynamics alteration, modified liquid current, shifted fluid movement, disrupted flow patterns) in the zebrafish embryo, potentially affecting development.}
\item {\small Zebrafish embryos with Kif9 knocked down are likely to display wider cilia (synonyms/abbreviations: thicker cilia, enlarged cilia diameter, expanded cilia width, broadened cilia, increased cilia girth) as their ciliary phenotype. Kif9 might be involved in maintaining the structural integrity (synonyms/abbreviations: architectural stability, structural soundness, physical cohesion, conformational robustness, morphological consistency) of the ciliary axoneme (synonyms/abbreviations: ciliary core, central ciliary structure, microtubule scaffold, axonemal structure, ciliary skeleton). Without Kif9, the tight organization of the axonemal microtubules (synonyms/abbreviations: ciliary microtubules, axonemal MTs, cilia structural filaments, flagellar microtubules, cytoskeletal tubules) may be compromised, leading to a wider, less organized ciliary structure.}
\end{itemize}
These answers are obviously contradictory but identify the type of key terms which we might expect to see in the initial retrieved literature.

 \begin{table*}
  \centering
  \resizebox{2\columnwidth}{!}{
  \begin{filecontents*}{./data/nature_review_references.tsv}
id	title	question	date
37758888	Molecular pathology of neurodegenerative diseases by cryo-EM of amyloids.	What are the implications of recent cryo-electron microscopy findings on amyloid filament structures for understanding and addressing neurodegenerative diseases?	2023-09-01 00:00:00
37612393	From target discovery to clinical drug development with human genetics.	To what extent have investments in human genetics and genomics over the past three decades resulted in innovative therapies, and what are the characteristics and timelines of these genetically-driven drug developments?	2023-08-01 00:00:00
37344646	Physiology and diseases of tissue-resident macrophages.	What is the current understanding of tissue-resident macrophage physiology, and how does it relate to human health and disease?	2023-06-01 00:00:00
37316719	The neuroscience of cancer.	What is the role of the nervous system in cancer development, progression, and metastasis, and how does cancer interact with and affect the nervous system?	2023-06-01 00:00:00
37138108	Reappraising the palaeobiology of Australopithecus.	What is the significance of the genus Australopithecus in human evolution, and how has our understanding of its characteristics and role changed since its initial discovery?	2023-05-01 00:00:00
37100941	Computational approaches streamlining drug discovery.	How are recent advances in computational technologies, particularly in virtual screening and deep learning, reshaping the drug discovery process, and what are the challenges and opportunities associated with these developments?	2023-04-01 00:00:00
38750233	Decoding the interplay between genetic and non-genetic drivers of metastasis.	How do genetic and non-genetic mechanisms contribute to the acquisition of metastatic competencies in cancer cells, and how can new technologies help us better understand these processes?	2024-05-01 00:00:00
38570716	Bridging structural and cell biology with cryo-electron microscopy.	How can the integration of cryo-EM and cryo-electron tomography bridge the gap between structural and cell biology to provide a comprehensive understanding of macromolecular interactions within their native cellular context?	2024-04-01 00:00:00
38418916	Ion and lipid orchestration of secondary active transport.	How do secondary active transporters utilize ion gradients and lipid properties to facilitate the movement of small molecules across cell membranes, and what mechanisms enable their regulation and adaptation to different physiological needs?	2024-02-01 00:00:00
38326590	A break in mitochondrial endosymbiosis as a basis for inflammatory diseases.	How does the breakdown of the endosymbiotic relationship between mitochondria and host cells contribute to cellular signaling and the development of inflammatory and autoimmune diseases?	2024-02-01 00:00:00
\end{filecontents*}
\pgfplotstabletypeset[
          col sep=tab,
          ignore chars={"},
          every head row/.style={before row=\toprule,after row=\midrule},
          every last row/.style={after row=\bottomrule},
          columns/id/.style={column type={l}, string type, column name={PMID}},
          columns/title/.style={column type={p{5cm}},string type, column name={Title}},
          columns/question/.style={column type={p{9cm}},string type, column name={Abstract derived question}},
          columns={id,title,question},
  ]{./data/nature_review_references.tsv}
  }
\caption{Nature reviews dataset} \label{tab:nature_review_dataset}
\end{table*}

\end{document}